\documentstyle[12pt,moriond,epsf1990]{article}
\def \physrep#1#2  {{\em Phys. Rep. \/} {\bf #1}, {#2}}
\def\.{{\cdot}}

\def\centreline{\centerline}

\def\fgaufluc{
\begin{figure}
\centering 
\centreline{\epsfxsize=14cm
\epsfbox[55 44 718 433]{"flucGampMor.ps"}
}

\caption{
\label{f-gaufluc} 
The left panel shows 
perturbations, with amplitudes drawn from a Gaussian distribution, 
shown in the right panel. The perturbations are shown as 
overdensity $\,\delta \equiv (\rho - \overline{\rho})/\overline{\rho}\;$ 
as a function of distance, scaled to 
a fixed wavelength (arbitrary units) and shifted to a fixed phase. 
The Gaussian probability density function 
and the perturbations are shown 
in units of r.m.s., i.e., ``1-$\sigma$'' fluctuation amplitudes 
(vertical axes).  
Amplitudes less than $0\.25$-$\sigma$ are shown in bold. If 
the collapse epoch of 1-$\sigma$ perturbations is 
$z_{\mbox{\rm \small coll}}=3,$ 
then $0\.25$-$\sigma$ perturbations should collapse at 
$z_{\mbox{\rm \small coll}}=0,$ 
(for $\Omega_0=1, \lambda_0=0$), forming ``type (vi)'' dwarfs.
}
\end{figure}
} 

\begin{document}
\heading{DWARF GALAXIES FROM \\COSMOLOGY}

\author{B.F. Roukema $^{1,2}$}
{$^{1}$ Observatoire de Strasbourg, Strasbourg, France.} 
{$^{2}$ Centrum Astronomiczne im. Mikolaja Kopernika, Warsaw, Poland.}

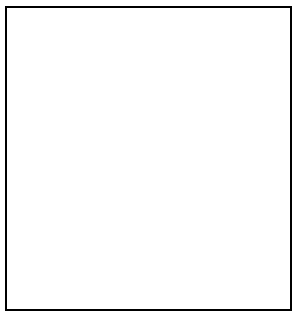
\begin{moriondabstract}
A local property of standard cosmology 
--- growth of primordial perturbations ---
implies at least six different
formation mechanisms of dwarf galaxies. 
A global property --- the topology 
of the Universe --- may enable direct observational study of
the aging of individual galaxies.

 Dwarfs (by mass) can form by star formation 
in low mass dark matter haloes, which generally collapse
early in hierarchical galaxy formation scenarios. However, 
several other formation paths exist, including that 
of low redshift, recently collapsed, low mass dark haloes 
containing low metallicity stellar populations.
The latter possibility is purely a property of gravitational 
collapse --- no special mechanisms to retard star formation
are required. Dwarfs by ``nurture'', as opposed to ``nature'',
are also possible.
\end{moriondabstract}

\section{Introduction}

In principle, we are here to talk about ``Dwarf Galaxies and Cosmology''.

Dwarf galaxies can be defined either as galaxies of low luminosity or
as galaxies of low mass. ``Cosmology'', as an observational science,
should be taken to mean the study (a) of global properties of the
Universe, to the extent that properties appear to be global, or (b) of
local properties of the Universe which are significantly affected by
global properties.

The formation of galaxies and large scale structure from primordial
fluctuations in the early Universe is a local property, 
i.e., an example of (b), while 
the study of the topology of the Universe (e.g., 
\cite{LaLu95}, \cite{Rouk96}) is a global property, i.e., an example of (a). 

Within a decade, both the curvature and the topology parameters 
of the Universe are hoped to be known 
to within a few percent accuracy to the limits
possible within the observable horizon. Once these are
known, and if the size of the Universe is small enough,
then for a given dwarf galaxy, multiple topological images 
of the same dwarf galaxy at different epochs (probably
by several Gyr) will be observable at predicted positions.

So, statistical studies of dwarf galaxy evolution
may be complemented by that of 
the studies of individual dwarf galaxy aging.

\section{Hierarchical galaxy formation and dwarf galaxies}

The most popular scenarios of galaxy formation 
from small density perturbations
within a cosmological context (e.g., \cite{Silk77,ReesO78})
are termed ``hierarchical'', since small length scale 
(i.e., small mass scale) perturbations
generally collapse first, and later merge together to form dark matter
haloes of successively higher and higher masses. 
Subsequent to 
gravitational collapse, cooling processes in the baryonic components
enable star formation. 

\subsection{Obvious dwarfs}

Dwarf galaxies defined either by mass or by luminosity can 
therefore be expected in three categories: 
\begin{list}{(\roman{enumi})}{\usecounter{enumi}}
\item galaxies in low mass haloes, at high $z,$ (dwarfs by mass);
\item galaxies formed in low mass haloes at high $z,$ which have escaped
merging and so are seen at low $z,$ (dwarfs by mass); and
\item galaxies in high mass haloes at low $z,$ which 
have had much less
star formation than other galaxies in high mass
haloes (dwarfs by luminosity).
\end{list}

Dwarfs (i) are likely to be difficult to observe, in particular
because of (bolometric) surface brightness dimming of $(1+z)^4$ 
(cf. \cite{Ferg,Pett}).

Dwarfs (ii) are probably the main observational target of this 
meeting, although some of them are detected in studies intended 
for the study of the high $z$ galaxies (e.g., the HDF \cite{Ferg}).

Dwarfs (iii), i.e., defined by luminosity, include dwarf spheroidals such as
DDO154 (e.g., \cite{PurCar96}) and 
high mass, low surface brightness galaxies (though because of the 
high mass, the term ``dwarf'' tends not to be preferred for these).

Since dwarfs of types (ii) and (iii) are those observed, these
are of immediate interest for detailed hierarchical galaxy formation
models. Since relative isolation is probably necessary to avoid either
merging of dark matter haloes [for dwarfs (ii)] 
or to avoid interaction-induced
star formation [for dwarfs (iii)], simulations based on 
statistical analytical approximations
(e.g., \cite{KW93}) may be sufficient to correctly model dwarf
galaxy formation and evolution. 

\subsection{Less obvious dwarfs}

However, there remain some less obvious categories of dwarfs (by mass)
expected in a hierarchical model, at least the first two for which the
modelling requires the inclusion of non-linear effects absent in
``semi-analytical models'':

\begin{list}{(\roman{enumi})}{\usecounter{enumi}}
\addtocounter{enumi}{3}
\item galaxies {\em formed} in high mass haloes at low $z,$ 
with a ``normal'' star formation rate, but which are stripped of 
both dark and luminous matter by interactions with other galaxies
or the inter-galactic medium inside of a cluster; 
\item galaxies formed in the tails of major galaxy mergers; and
\item galaxies in low mass haloes formed at {\em low} $z.$ 
\end{list}

In principle, the non-linear effects are included 
in full-scale hydro-gravitational $N$-body simulations
(e.g., \cite{WeinbHK97}). Alternatively, pure gravity $N$-body 
simulations can be combined with analytical formulae using the
method of \cite{RPQR97}, which is being updated into a user-friendly
software package, {\sc ArFus} (``arbres de fusions'', 
\cite{RNDBGM98}). 

If dwarfs (iv) exist, then they have probably already been observed 
and need to be distinguished from dwarfs (ii).

At least some dwarfs (v) exist (\cite{Duc}).

At least some dwarfs (vi) also exist (\cite{Thuan, Izotov}), 
and these are {\em not}
in contradiction with the standard hierarchical galaxy formation
scenario.

\subsection{Low amplitude fluctuations: dwarfs of ``type (vi)''}

Fig.~11 of \cite{RPQR97} shows that the formation of dwarfs (vi)
is apparent in $N$-body derived 
merging history tree simulations of galaxy formation: 
the ``merging histories'' of low $z$, low mass galaxy haloes 
show recent birth and little or no merging.

The explanation for this provides a useful reminder about
``primordial Gaussian fluctuations''. 

``Gaussian'' does not refer to the shape of the 
fluctuations; it refers to the probability distribution of
the {\em amplitudes} of the fluctuations. 
Fig.~\ref{f-gaufluc} shows 
an example of some fluctuations, at a fixed length scale
and fixed phase, drawn from a Gaussian distribution
of amplitudes. 

\fgaufluc

In fact, the most common fluctuations are of zero amplitude 
and so do not collapse to form structure! Of course, the 
majority of fluctuations have amplitudes $A$ with, say,  
$|A| > 0\.25,$ in units of the r.m.s. amplitude $\sigma.$ 
So a majority of small length scale perturbations, whose amplitudes
have $|A| \sim 1,$ generate the low mass haloes which
should collapse early and provide the low mass dwarfs [types
(i) and (ii)].

Nevertheless, some small length scale fluctuations have low, but
non-zero amplitudes. These fluctuations, according to ``linear
theory'', should only collapse at recent epochs. 
If $z_{\mbox{\rm \small coll}}$ is the collapse epoch of 
haloes of a given mass generated from 1-$\sigma$ fluctuations,
then fluctuations of $[1/(1+z_{\mbox{\rm \small coll}})]$-$\sigma$
amplitude will only just be collapsing now,
forming dwarfs of type (vi)
 [for $\Omega_0=1,
\lambda_0=0$, in which case amplitude growth is 
$\delta \propto 1/(1+z)$].

Of course, some of the small length scale fluctuations expected
to collapse at the present may be inside of large length scale
fluctuations, i.e., they merge before they have
a chance to collapse, and end up in a higher mass halo. 

Alternatively,
the small length scale fluctuations may contain 
fluctuations on yet smaller scales
(above the recombination epoch Jeans mass), which mostly collapse earlier.
In this case, the low $z$ newly formed halo would be partly composed
of these smaller haloes which could already be metal-enriched via
star formation. Hence, type (vi) dwarfs should not be too much greater 
than the Jeans mass if they are to contain zero metallicity matter 
and no old stars.

The existence of some 
low $z,$ low $Z,$ low mass galaxies is not 
surprising in hierarchical galaxy formation. This property does not require
any special assumptions regarding star formation --- it is simply
gravitational.

Moreover, precise statistics on the numbers, masses and 
metallicities of such galaxies, combined with the recombination epoch Jeans
mass, might provide an independent method of constraining the
cosmological metric parameters, $\Omega_0$ and $\lambda_0,$ in the
sense that a high number 
density of low $z$-$Z$-$M$ galaxies would provide a constraint 
against a low density of the Universe
(since perturbation 
growth slows near the present in a low density universe).

\acknowledgements{This work has been supported by the 
Observatoire de Strasbourg, CNRS and by the 
Polish Council for Scientific Research Grant
KBN 2 P03D 008 13.}


%
\vfill
\end{document}